\documentclass[10pt,nobibnotes,preprint,injp,nofootinbib,showpacs]{article}
\usepackage{latexsym}
\usepackage{amssymb}
\usepackage{amsmath}
\usepackage{amscd}
\usepackage{amsthm}
\usepackage{times}
\usepackage{mathrsfs}
\usepackage[left=2cm,top=2.5cm,right=2.5cm,bottom=1.5cm]{geometry}
\usepackage[dvips]{graphicx}
\usepackage{bm}
\begin{document}
\begin{center}
\large{\bf{A model study on atom-atom interactions with large scattering length in quasi-two  dimensional traps}} \\
\vspace{10mm}
\normalsize{ P Goswami$^1$, A Rakshit$^1$ and B Deb$^{1,2}$ } \\
\vspace{5mm}
\normalsize{$^1$ Department of Materials Science $^2$ Raman Centre for Atomic, Molecular and Optical Sciences, \\
Indian Association for the Cultivation of Science,
Jadavpur, Kolkata 700032. INDIA.}
\end{center}
\begin{abstract}
We carry out a model study on two-atom interactions and bound states  in quasi-two  dimensional traps.
 The interactions  are modeled by two-parameter potentials with parameters being the range $r_0$ 
 and the $s$-wave scattering length $a_s$. We show that one can make use of two forms of finite-range model 
 potentials: one  for $a_s > 0$ and the other for $a_s < 0$. Both potentials reduce to same form in the limits
$a_s \rightarrow \pm \infty $.  We investigate into the dependence of the binding energies and 
 the wave functions of two-atom trap-bound states  on $a_s$ and $r_0$. In particular, we study the effects
 of $a_s$ ranging from large negative to large positive values on the bound state properties.
 Our results show that long-range interactions with infinite scattering length 
 significantly alter  the ground-state energy of the two atoms in a quasi-two or two dimensional trap. In contrast, short-range interactions 
can not significantly change the ground-state energy of two atoms in a 2D harmonic trap.  
\end{abstract}


\def\be{\begin{equation}}
\def\ee{\end{equation}}
\def\bea{\begin{eqnarray}}
\def\eea{\end{eqnarray}}
\def\zbf#1{{\bf {#1}}}
\def\bfm#1{\mbox{\boldmath $#1$}}
\def\hf{\frac{1}{2}}
\smallskip

PACS number {34.10.+x; 37.10.Gh; 94.30.Hn}

\section{Introduction}

It is well established that the many-body properties of one or two dimensional systems \cite{Many-Particle Physics}
are qualitatively different from those of three dimensional ones. 
Generally, interacting systems in low dimensions exhibit intriguing properties,  for example,  
exotic phenomena such as quantum Hall effect and high temperature superconductivity occur in electrons in two dimensions (2D). 
Over the last three decades, low dimensional phenomena have been extensively 
studied in condensed matter systems. With the recent  advent of ultracold atoms  in highly anisotropic traps, new 
perspectives of low dimensional physics with trapped atoms have arisen \cite{jphys:petrov:2004}. 
An ensemble of harmonically trapped noninteracting  ultracold atoms
becomes kinematically two dimensional in $x$ and $y$ directions (one dimensional in $z$ direction) when the temperature and the chemical potential 
of the ensemble are much lower than the harmonic trapping frequency in the tightly confining $z$ direction (frequencies in $x$ and $y$ directions). 
In such a 2D (1D)  trap, the atoms essentially occupy the ground states (state) 
of the harmonic oscillators in the tightly confining  directions (direction). Then, the deviations from the purely 2D (1D) 
physics are expected to arise only from interatomic interactions. The effects of tight trapping or lower dimensions   
on Bose-Einstein condensates (BEC) have attracted a great deal of research interests, 
 both theoretically [3-9]
and experimentally [10-13]
ever since the first realization 
 of atomic Bose-Einstein condensation in 1995.

One  unique advantage for research with  ultracold atoms is that  one can alter
 interactions between the atoms over a wide range by a magnetic Feshbach resonance 
[14-16] This provides  
an opportunity for exploring many-body physics with tunable interactions \cite{rmp:2008:bloch}.  
Though, most of the recent experimental works on ultracold atomic gases with 
tunable $s$-wave scattering length  are carried out 
in 3D traps, it is possible to study tunable two-body interactions in  highly anisotropic or 
lower dimensional traps \cite{pra:2002:bolda}. 
The effects of confinement due to tight anisotropic trapping on atom-atom cold collision 
are an interesting topic of research in cold atom science. 
Confinement induced resonances  in
 quasi-one dimension due to strong confinement along transverse directions  for any value
 of 3D scattering length have been discussed theoretically by Olshanii \cite{prl:1998:olshanii}. It has been experimentally demonstrated
 by Moritz et al. \cite{prl:2005:moritz} that, cold atoms trapped in quasi-1D traps can form  
 confinement  induced dimers.  Petrov and Shlyapnikov \cite{pra:2001:petrov} have discussed quasi-2D and 3D regimes of atom-
atom scattering in a trap that is tightly confined in axial direction.

\begin{figure}
 \centering
 \includegraphics[width=3.5in]{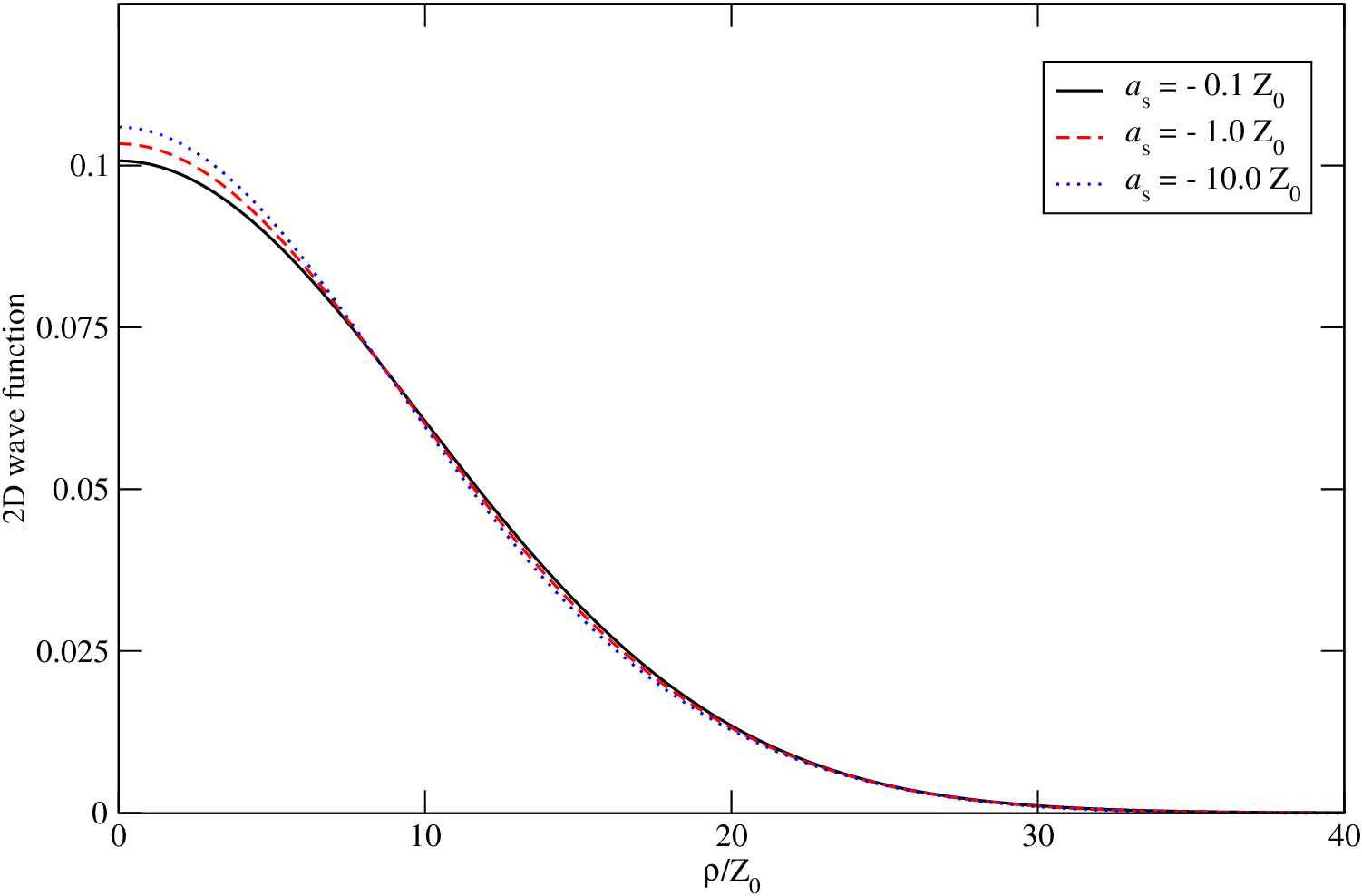}
\caption{ Quasi-2D ground state wave function in unit of $\rho_0^{-1}$ for  $a_s = -0.1 Z_0$ (solid line), 
$a_s = -1.0 Z_0$ (dashed lines )
and  $a_s = -10.0 Z_0$ (dotted lines)for the 
fixed parameters $\eta =100$ and $r_0/Z_0 = 1$}
 \label{fig:5}
\end{figure}

Usually, two-body interaction in dilute atomic gases  at low temperatures is modeled with 
a zero-range contact potential.  In such an approach, 
the actual two-body interaction is replaced by a delta-type pseudo-potential 
\bea 
V_{\rm{pseu}} = \frac{4 \pi \hbar^2 a_s}{2\mu} \delta (\mathbf{r})
\label{pseudo}
\eea 
where 
$a_s$ represents the  energy-independent $s$-wave scattering length, $\mu$ is the reduced mass of two colliding atoms and 
$\mathbf{r}$ denotes the separation between the two particles. This is valid for a class of two-body potentials which
have finite  $a_s$ in the zero energy limit. 
The essential idea behind this approach is to deal with a simple potential that is 
capable of reproducing the total two-body elastic scattering amplitude in the limit of the collision energy tending to zero. 
 The $s$-wave scattering amplitude $f_k$ for collision
wave number $k$ is related to the $s$-wave scattering phase shift $\delta (k)$ by 
\bea 
f_k = \frac{1}{k \cot \delta_{k} - i k} 
\eea 
and $a_s$ is related to $\delta_k$ by the well-known  Bethe's expansion formula
\cite{bethe}
\begin{equation}
 \lim_{k\rightarrow 0} k \cot[\delta(k)] = -\frac{1}{a_s} + \frac{1}{2} r_0 k^2 + ...
 \label{expan}
\end{equation}
where  $r_0$ is the effective range.
Near a scattering resonance,  $a_s$ diverges and so the potential, shown by expression (\ref{pseudo}), becomes ill-defined. 
This means that  
the delta function potential can not rigorously describe the phenomena that occur near a scattering resonance. However, 
an improved  
treatment of resonant phenomena can be done in terms of the regularised delta function potential
derived by Lee {\it et al.} \cite{PHYSICAL REVIEW:Huang:1957}. Busch {\it et al.} \cite{foundphys:1998:busch} 
have obtained an exact solution and bound-state energy spectrum for two ultracold atoms interacting via zero-range regularized delta potential in an isotropic 3D harmonic 
trap.  Idziaszek and Calarco \cite{pra:2005:idziaszek} have
found an exact solution of two interacting particles 
in an axially symmetric trap within pseudopotential approximation. 
Tiesinga {\it et al.} \cite{pra:2000:tiesinga} have argued that the applications of the exact solutions obtained in Ref.\cite{foundphys:1998:busch} are 
limited to the sufficiently weak traps the width of which is much larger than $|a_s|$. To circumvent this limitation,  
Bolda and coworkers \cite{pra:2002:bolda} have used an energy-dependent scattering length 
in a regularised pseudopotential and developed a self-consistent method to calculate two-atom bound state  
in an isotropic trap, and to study  $s$-wave collisions in an optical lattice with quasi-one and 
two dimensional harmonic confinement \cite{pra:2003:bolda}. 
Peach {\it et al.} \cite{pra:2004:peach} have theoretically studied ultracold collisions between metastable helium atoms in 
tight harmonic traps using energy-dependent scattering length based self-consistent as well as 
quantum defect theoretic numerical integration methods. 

The  purpose of our investigation is  
to understand how a finite-range resonant two-body interaction  can affect  
two-body bound states in low dimensional traps. To this end, 
we use a class of finite-range model potentials 
which do not diverge as $a_s \rightarrow \pm \infty$ and  do not  require any regularisation. To treat the effects of large 
scattering length in a tightly confined trap, our model potentials does not require any assumption of an energy-dependent scattering length.  
These model potentials are based on the expansion  (\ref{expan}) and derivable by the method of Gelfand and Levitan \cite{Levitan:1951}. 
We show that there exists two such finite-range 
model potentials - one for $a_s > 0$ and the other for $a_s < 0$. In the limits $a_s \rightarrow \pm \infty$, 
both potentials reduce to the same form. The finite-range model potential for  $a_s =  - \infty $ is well-known and used earlier  
by Carson {\it et al.}  \cite{prl:2003:Carlson} 
for quantum Monte Carlo simulation of a homogeneous 
superfluid Fermi gas, and by also  Shea {\it et al.} \cite{ajp:2008:Bhaduri}
to study the energy spectrum of two interacting cold atoms in an isotropic harmonic trap. 
Following the 
work of Jost and Kohn \cite{kohn:1953}, we introduce an analytical form of 
the potential for $a_s >  0$ that can smoothly match with the other potential for $a_s \rightarrow - \infty $ in the limit 
$a_s \rightarrow + \infty$. This means that these two potentials can account for the entire regime of low-energy interactions from 
large positive to the large negative scattering length for any arbitrary range of interactions.

\begin{table}
\caption{\label{abrefs} Ground-state energy eigenvalue $E_{g}^{{\rm 3D}}$ (in unit of $\hbar\omega$) of 
two atoms  interacting with the model potentials with $a_s = \pm \infty$  in an isotropic 3D harmonic oscillator
 for different values of $r_0$ (in unit of $l_0 = \sqrt{\hbar/\mu\omega }$, where $\omega$ is the frequency of the 
 isotropic harmonic oscillator)
 applying Numerov method in one case and Hamiltonian matrix diagonalisation method in another case. }
\begin{center}
\begin{tabular}{lcc} \hline
$r_0$                  & $E_{g}^{{\rm 3D}}$ (Numerov)   & $E_{g}^{{\rm 3D}}$ (Matrix)                                                              \\ \hline 
10.0                    & 1.4622   &1.4622                                  \\
6.0                   & 1.4045    & 1.4045                                 \\
4.0                   &1.3162    & 1.3163                                  \\
2.0                    &1.0742     & 1.0747                                  \\
1.0                    &0.82220      & 0.8507                                      \\     \hline
\end{tabular}
\end{center}
\end{table}

Here  we study the effects of the scattering length 
and the range of the potentials on the two-atom bound-state properties in  quasi two-dimensional (quasi-2D) traps. 
Our results illustrate that, 
as the positive scattering length increases,
the probability amplitude for finding the two  particles at the  centre of the trap
decreases, while that in the case of negative scattering length increases. The ground-state energy 
of  two-particle bound state in the trap approach the same value for both limits $a_s \rightarrow \pm \infty$. 
Our results further demonstrate that the long-ranged interactions most significantly affect two-atom bound states in 
quasi-2D  while short or zero-ranged interactions have rather small effects on such bound states.

\begin{table}
\caption{\label{abrefs}Six low-lying energy eigenvalues $E_{\nu,m}$ (in unit of $\hbar\omega_{\rho}$) of two-atom  
bound states in quasi-2D  harmonic oscillator traps  ( $\eta=100$ and  $\eta=1000$ )
and  different values of $a_s$ (in unit of $Z_0$) keeping  $r_0 = 1Z_0$. }
\begin{center}
\begin{tabular}{lccccccc} \hline
$\eta$                                       &$a_s$                  &   $E_{0,0}$                 & $E_{1,0}$     & $E_{2,0}$ &  $E_{0,1}$               & $E_{1,1}$              & $E_{2,1}$  \\ \hline 
100                   & 4                 &0.9055     & 2.9125       &  4.9173  &1.999380       &3.998790       &5.998230 \\
100                   & 10                 & 0.9153     & 2.9209        &4.9247   &1.999530      &3.999080       &5.998640  \\
100                   & $\infty$            &0.9201       &2.9250        &4.9285  &1.999590     &3.999264       &5.998910  \\     \hline 
1000                  &4                 & 0.9715         &2.9721        &4.9725   & 1.999999      &3.999960       &5.999940    \\ 
1000                  &10.0                  & 0.9743      &2.9748        &4.9751    &1.999980      &3.999970       &5.999950    \\ 
1000                  &$\infty$             &0.9757        &2.9761         &4.9764    & 1.999980     &3.999970       &5.999960     \\     \hline
100                   & -0.1                 &0.9918       &2.9918         &4.9919  &1.999995      & 3.999989        &5.999980   \\
100                   & -1.0                 &0.9623       &2.9634         &4.9642   &1.999899       &3.999800         &5.999700        \\
100                   & -10                  &0.9321        &2.9357         &4.9383   &1.999695       &3.999398       &5.999100      \\
100                   & -$\infty$            &0.9245        &2.9289         &4.9320   & 1.999627      &3.999264       &5.998910      \\   \hline
1000                  &-0.1                  &0.9974       &2.9974         &4.9974    & 1.999999      &3.999999       &5.999999       \\
1000                  &-1.0                  &0.9880        &2.9880          &4.9880   &1.999996       &3.999993       &5.999990       \\ 
1000                  &-10.0                 &0.9782        &2.9785         &4.9788    &1.999989       &3.999979       &5.999960        \\
1000                  &-$\infty$             &0.9757        &2.9761         &4.9765    &1.999987       &3.999974       &5.999960        \\   \hline
\end{tabular}
\end{center}
\end{table}

\section{Model potentials}

The model interaction potentials,  we consider,  are two-parameter potentials. 
The two  parameters are the range $r_0$ of the potential and the 
$s$-wave scattering length $a_s$. These potentials are derived making use of the effective range expansion of Eq. (\ref{expan}). 
The procedure 
for deriving a finite-ranged effective 
potential from the experimental data of phase shift $\delta_0(E)$ 
was first demonstrated long ago by Fr\"{o}berg \cite{Ark Mat Astr Phys:1948}. This 
was followed by the work of Gel'fand and Levitan \cite{Levitan:1951} who gave the mathematical method
for the derivation of finite-range model potentials. This method was then used by a number of workers in 
deriving model potentials for various physical systems and  parameter regimes.

For an appropriate form of a finite-range potential for negative $a_s$, we use the potential introduced by 
Jost and Kohn \cite{PHYSICAL REVIEW:Kohn:1952}.
This has the form
\bea 
V_{-}(r) = - \frac{4 \hbar^2}{ \mu r_0^2}  \frac { \alpha \beta^2 \exp(-2 \beta r/r_0) }{ [ \alpha + \exp(-2 \beta r/r_0) ]^2} 
\label{negative}
\eea
where $\alpha = \sqrt{1 - 2 r_0/a_s}$,  $\beta =  1 + \alpha $ and $\mu$ is the reduced mass. 
This potential is valid for $|\delta_0(E)| < \pi/2$ in the limit $E \rightarrow 0$.

For positive $a_s$, we make use of the  three-parameter  potential derived again by Jost 
and Kohn \cite{kohn:1953} from an adaptation of 
the mathematical method of Gel'fand and Levitan \cite{Levitan:1951}. 
Among the three parameters,  two are $r_0$ and $a_s$,  the third one `$\lambda$' \cite{kohn:1953})  is
related to the binding energy of a bound state that the potential may support. 
If we  use the particular choice $ \lambda = -\sqrt{1-2r_0/a_s}$ with $a_s > 2r_0$, 
 then the potential given by Eq. (2.29) of  \cite{kohn:1953} reduces to a 
two-parameter potential of the form 
\bea 
V_{+}(r)=   - \frac{4 \hbar^2}{ \mu r_0^2}  \frac { \alpha \beta^2 \exp(-2 \beta r/r_0) }{ [ 1 + \alpha \exp(-2 \beta r/r_0) ]^2}.  
\label{positive}
\eea
This choice of  $\lambda$'corresponds to the binding energy $E_{\rm{bin}} \simeq \hbar^2/(2 \mu a_s^2)$ for $2 r_0/a_s <\!<1$. 
If a potential supports a bound state, $a_s$ is positive. However, positivity of $a_s$ is not sufficient for a potential to support 
a bound state. 
As discussed in  \cite{kohn:1953}, the experimental data on phase shift and the range of a two-body interaction are not enough 
to construct a model potential that can support a bound state. In 
order to construct such a potential it is necessary to add another term making use of the bound state wave function.  This results in a 
four-parameter potential, the fourth parameter being the normalization constant of the wave function. Here we do not consider such 
bound-state supporting interaction potentials.

In the limit $a_s \rightarrow \pm \infty$,
both the potentials given by Eqs. (\ref{negative}) and (\ref{positive}) reduce to the same form
\bea 
V_{\infty} =   -\frac{4\hbar^2}{\mu r_0^2 \cosh^2(2r/r_0)}
\label{infty}
\eea
Thus the two potentials of Eqs. (\ref{negative}) and (\ref{positive}) 
smoothly connect to the resonance point as $a_s$ is varied through $\pm \infty$. 

\begin{figure}
 \centering
 \includegraphics[width=3.5in]{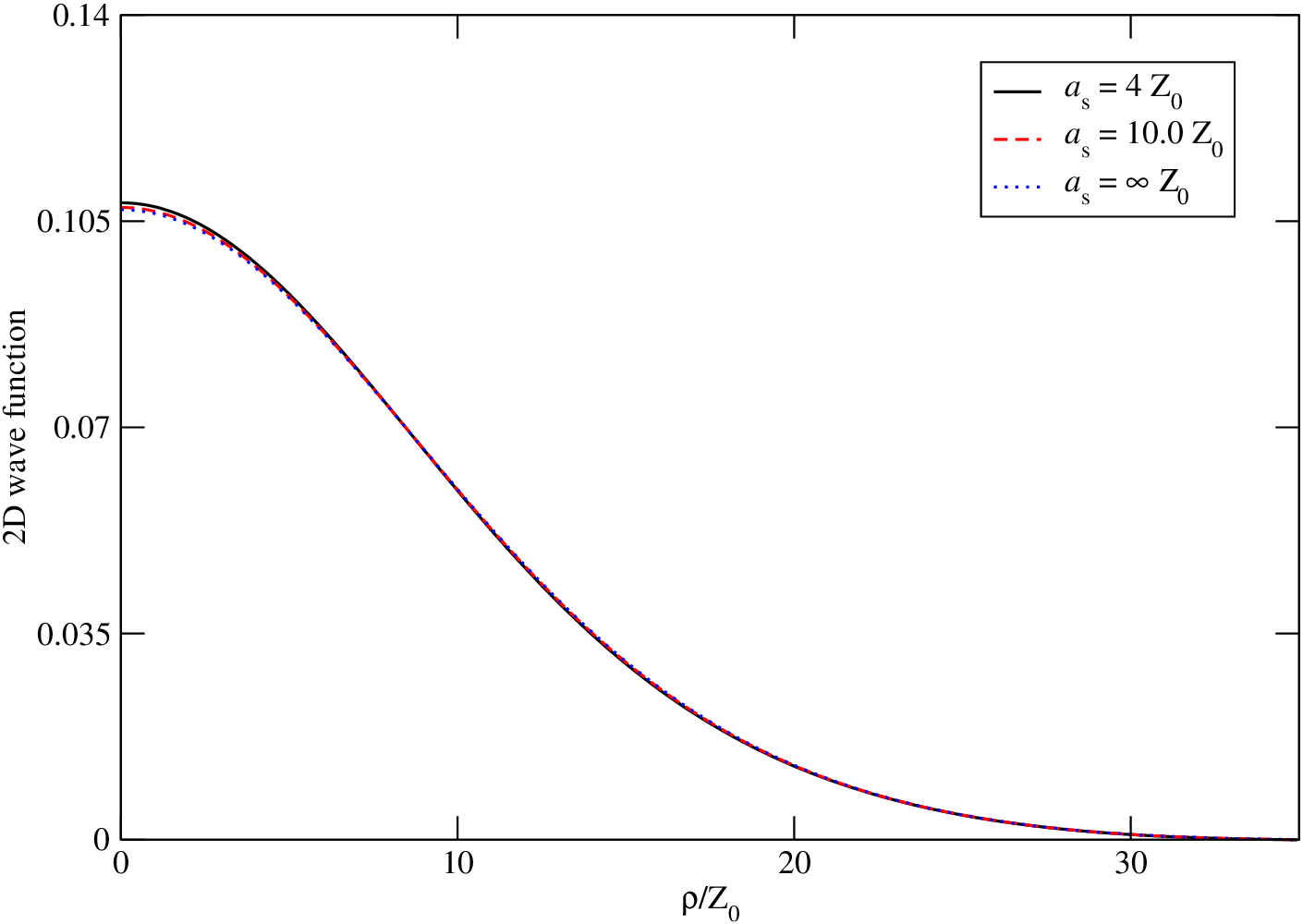}
\caption{ Quasi-2D ground state wave function in unit of $\rho_0^{-1}$ for  $a_s = 4 Z_0$ (solid line)  $a_s = 10.0 Z_0$ (dashed red lines )
and  $a_s = \infty$ (dotted  lines). The other parameters remain same as in Fig. \ref{fig:5}.}
 \label{fig:6}
\end{figure}

\section{Bound states in a harmonic trap}

The question is how to treat resonant interaction in a highly anisotropic or low dimensional 
trap. In the case of atoms inside traps, two-body interaction can not be  described from
the traditional scattering point of view. Because, atoms are bound in a trap, and the interaction between two atoms inside 
the trap can change the properties of trapping states of atoms. In case of two atoms in a harmonic trap, the centre-of-mass and the relative
motions between the two atoms become separable. As a result,  only the relative motion  will be affected by an isotropic interaction. 
The effects of 
interactions will then be manifested through the changes in the properties of trap-induced two-atom bound states.  
In this context, it is worth noting that, very recently variable interatomic interaction induced shifts in the bound-state energy of 
two ultracold atoms in the microtraps of a 3D optical lattice have been used to experimentally demonstrate Feshbach resonance between 
excited- and ground-state atoms \cite{Takahashi:2013:prl}.   In case of highly 
anisotropic harmonic traps,  lowering of spatial dimensions is possible. Then it is necessary to discuss the effects of low dimensionality
on the bound states. 

A  harmonic trap is anisotropic if trapping frequencies in all three directions are not same.  
Let us consider axially symmetric case i.e $\omega_x = \omega_y = \omega_{\rho} \ne \omega_z$.  
 If $\omega_{\rho}$ $>>$ $\omega_z$ i.e. if the trapping 
frequency $\omega_{\rho}$ in radial direction is much greater than that in axial direction, we get a cigar-shaped one dimensional (1D) trap.
On the other hand, if $\omega_z
>> \omega_{\rho} $ then we have pancake like quasi-two  dimensional trap.

Since the relative and centre-of-mass motions between two atoms  in a harmonic trap are separable, and the 
model interaction potentials we consider are isotropic, we henceforth consider only the relative motion between two atoms 
in an axially symmetric  harmonic trap. In case of a many-particle system in a harmonic trap, 
the system  effectively becomes two dimensional if the chemical potential and thermal energy are smaller than the energy gap in 
strongly confined axial direction. 
Our aim here is to elucidate  how finite-range interaction between a pair of atoms in a quasi-2D or 2D trap 
affects bound states between the atoms. Denoting the position coordinates of atom 1 and 2 by ($x_1,y_1,z_1$) 
and ($x_2,y_2,z_2$), respectively;  the cylindrical coordinates for the relative motion between the two atoms 
are given by $\rho = \sqrt{(x_1 - x_2)^2 + (y_1 - y_2)^2}$ and $z=z_1 - z_2$.  Schr\"{o}dinger equation for relative motion  
can then be written as 
\bea
&& \left [-\frac{\hbar^2}{2\mu} \left (\frac{d^2 }{d\rho^2} +\frac{d}{\rho d\rho} -\frac{|m|^2}{\rho^2} +\frac{d^2}{dz^2} \right ) + 
\frac{\mu}{2} \left (\omega_{\rho}^2\rho^2
   + \omega_z^2 z^2 \right) \right. \nonumber \\ 
 &+&  \left. V_{s}(\rho,z) \right ]\psi_{s}(\rho,z) =  E \psi_{s} (\rho,z)
 \label{eigen}
\eea
where the subscript $s$ stands for either `+' or `-', $\mu$ is the relative mass and $m$ is the magnetic quantum number.
Due to interaction, radial and axial modes can not be separated out. However,
in case of two noninteracting particles in an axially symmetric harmonic oscillator trap, the 
radial and axial modes. 
Let $\psi_{n_{\rho},m,{n_z}}^{0}(\rho,z)$ denote the wave function of relative motion between  a pair of noninteracting particles
in the trap, 
where  $n_{\rho}$ and $n_z$ are the 
radial and axial principal quantum numbers, respectively. This wave function 
is separable in axial and radial coordinates as 
\begin{equation}
\psi_{n_{\rho},m,{n_z}}^{0}(\rho,z) = R_{n_{\rho},m}^{0}(\rho) \times f_{n_z}^{0}(z) \\
\end{equation}
where
\bea
R_{n_{\rho},m}^{0}(\rho) = \left[\frac{n_{\rho}!}{\pi\Gamma(n_{\rho}+|m|+1)}\right]^{\frac{1}{2}}\frac{1}{\rho_{0}^{|m|+1}}\rho^{|m|} 
 \exp\left[{-\frac{\rho^2}{2 \rho_{0}^2}}\right]L_{n_{\rho}}^{|m|}\left(\frac{\rho^2}{\rho_{0}^2}\right) 
\eea 
and 
\bea
f_{n_z}^{0}(z) = \frac{\pi^{-\frac{1}{4}}}{\sqrt{2^{n_{z}}{n_z}!}}H_{n_z}\exp\left [-\frac{z^2}{2 Z_{0}^2} \right ]. 
\eea
$R_{n_{\rho},m}^{0}(\rho)$ and $f_{n_z}^{0}(z)$ are 2D and 1D harmonic oscillator wave functions, respectively. Here 
$L_{n_{\rho}}^{|m|}$ and $H_{n_z}$ are Laguerre and Hermite polynomials, respectively; 
$\rho_{0} = \sqrt{\frac{\hbar}{\mu\omega_{\rho}}}$ and $Z_0 = \sqrt{\frac{\hbar}{\mu\omega_{z}}}$ are the 2D and 1D harmonic
oscillator length scales, respectively. These two length scales are related by $\rho_0 = \sqrt{\eta} Z_0$.  
The energy $E = (2n_{\rho}+|m|+1)\hbar\omega_{\rho} +(n_z+1/2)\hbar\omega_z$ is the sum of 2D and 1D harmonic oscillator eigen energies. The aspect ratio of an axially symmetric trap is defined by 
$\eta = \omega_z/\omega_{\rho}$. 

\begin{figure}
\centering
\includegraphics[width=3.5in]{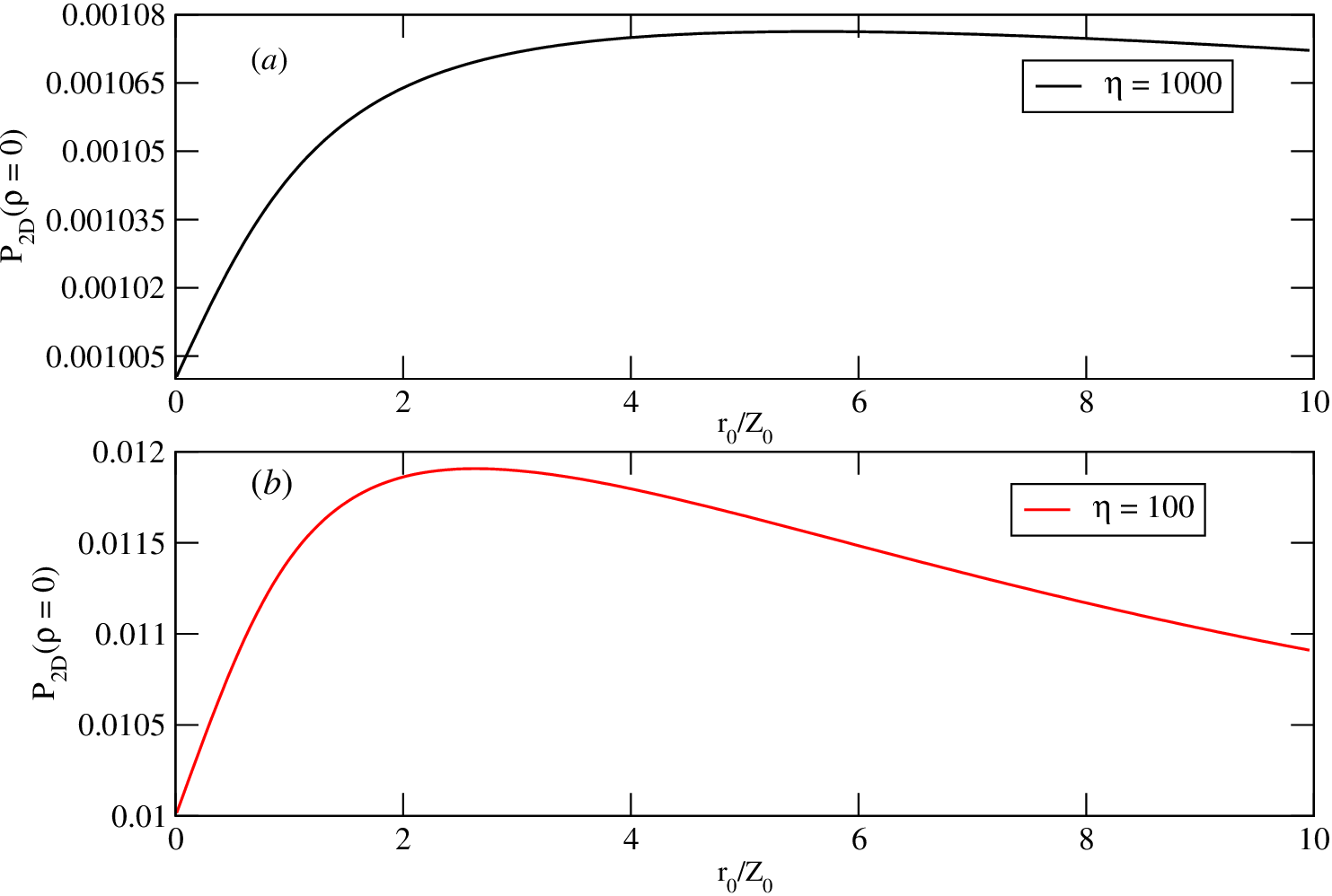}
\caption{The probability densities  $P_{2D}$  in unit of $\rho_0^{-2}$ 
at the trap centre ($\rho =0 $) versus $r_0/Z_0$ of $V_{\infty}$ for (a)  $\eta = 1000$  and
(b) $\eta = 100$. } 
 \label{fig:1}
\end{figure}

\subsection*{Quasi 2D or 2D regime}
We now return to the problem of two interacting atoms in an axially symmetric anisotropic potential. A quasi 2D or 2D regime 
may be reached when the two atoms are strongly confined in the axial direction i.e. ($\omega_x = \omega_y = \omega_{\rho} <<\omega_z$), 
so that the axial 
ground state is not affected much by the interaction.  
We call a trap quasi 2D if $\eta$ is relatively large and typically of the order of 100, for which the effects of axial motion at low energy is small. The 2D case is characterized  with 
$\eta$ of the order of or greater than 1000,  consistent with the recent experimental explorations \cite{prl:2011:kohl,nature:2012:kohl} of 
2D physics with trapped cold atoms with tight confinement along the axial direction. 
For 2D atom traps, typically value of $\omega_{\rho}$ is a few 
Hertz or of the order of 10 Hz while $\omega_z$ typically ranges between 10 to 100 kHz. This means that the 
harmonic oscillator radial length scale $\rho_0$  is in the micrometer regime,  while 
the axial length scale $Z_0$ is in the nanometer or even sub-nanometer regime.  Since 
$a_s$ of alkali atoms currently used for cold atom research is typically a few nanometer, the value $Z_0$ of a 2D trap
is comparable with $a_s$. 

Since the model 
interaction potentials are isotropic, the magnetic quantum number $m$ is a good quantum number.
Let $\psi_{\nu,m}^{(s)}$ denote an eigenfunction that satisfy  Eq. (\ref{eigen}), where $\nu$ is the principal quantum number 
of the interacting system. 
Expanding this wave function in terms of the 
wave functions of noninteracting one, we have 
\bea 
 \psi^{(s)}_{\nu,m} (\rho,z)  = \sum_{n_{\rho},n_{z}} c_{n_{\rho},n_z } \psi_{n_{\rho},m,n_z}^0 (\rho,z)
\eea 
Substituting $\psi_{s}(\rho,z)$ of equation (7) by $\psi^{(s)}_{\nu,m} (\rho,z)$, multiplying both sides by 
$\psi_{n_{\rho}^{\prime},m, n_z^{\prime}}^{0}(\rho,z)$ of Eq. (8)  
and then integrating over $z$ and $\rho$, we obtain 
\bea
(2n_{\rho}^{\prime} +|m|+1)c_{n_{\rho}^{\prime},n_z' } \hbar\omega_{\rho}  +   \sum_{n_{\rho}=1}^\infty \sum_{n_z} 
V_{s,n_{\rho}^{\prime} n_{\rho}; n_z' n_z }^{(m)} c_{n_{\rho}, n_z} = \tilde{E} c_{n_{\rho}^{\prime},n_z'}
\label{psi}
\eea 
where \bea  V_{s,n_{\rho}^{\prime} n_{\rho}; n_z',n_z }^{(m)} = \int \int d^2\rho d z 
R_{n_{\rho}^\prime,m}^0({\rho})  f_{n_z'}^0(z)  f_{n_z}^0 V_{s}(\rho,z) R_{n_{\rho},m}^0({\rho})\eea 
 and $\tilde{E} = E^{{\rm trap} } - \hbar \omega_z(n_{z}^{\prime} + 1)/2$, where  $E^{({\rm{trap}})}$ denotes an eigenenergy $E$ of 
 for the relative motion of the two interacting atoms in the anisotropic trap including both radial and axial modes.  
 Eq. (\ref{psi}) can be cast into a matrix form which can be  diagonalised 
 to evaluate the coefficients $c_{n_{\rho},n_z}$.

We first consider the ground state of our quasi 2D or 2D system with  $n_z = n_z' = 0$, that 
is, we assume that the atoms occupy only the ground state of their relative motion in the  axial direction. By solving 
Eq. (\ref{psi}), we obtain a 2D eigenfunction 
\bea 
\psi_{\nu,m}(\rho) =  \sum_{n_{\rho}} c_{n_{\rho},0 } R_{n_{\rho},m}^0 (\rho)
\eea 
with 2D energy eigenvalue $E_{\nu,m} = E^{{\rm trap}}_{\nu m} - \hbar \omega_z /2$. We then consider corrections to the ground state 
energies and wave function due to the finite probability of excitations to the higher levels of axial modes ($n_z \ne 0$). For 
calculating corrections to the ground state energy and wave function, we consider possible couplings 
of unperturbed states with $n_z = 0$ with the states with $n_z \ne 0$ due to the interaction.  Now, since the potential $V_s$ 
is an even function of $z$, the matrix element  $ V_{s,n_{\rho}^{\prime} n_{\rho}; n_z',n_z }^{(m)}$ will couple states 
with either even or odd $n_z$. Since we are primarily interested in the ground state energy and wave function,  we consider coupling
between even states only for our numerical work.
Let the minimum number of radial states required in order to get convergent results be $N_{\rho}$. 
Considering coupling between $n_z = 0$ and $n_z = 2$ and introducing  two vectors 
\bea  
 X_0 = \left ( \begin{array}{c}  c_{1 0} \\ c_{2 0} \\ \cdots \\
 c_{N_{\rho} 0} \end{array}  \right )
\eea 
and
\bea  
 X_2 = \left ( \begin{array}{c}  c_{1 2} \\ c_{2 2} \\ \cdots \\
 c_{N_{\rho} 2} \end{array}  \right )
\eea 
we can write the eigenvalue equation in the form 
\bea 
\left ( \begin{array}{c c}
         {\mathbf{H}}_{0} & {\mathbf{0}} \\
         {\mathbf{0}} & {\mathbf{H}}_{2} 
        \end{array} \right ) 
        \left ( \begin{array}{c}
                 X_0 \\
                 X_2 
                \end{array} \right ) 
                + \left ( \begin{array}{c c}
         {\mathbf{V}}_{0} & {\mathbf{V}}_{02} \\
         {\mathbf{V}}_{20} & {\mathbf{V}}_{2} 
        \end{array} \right ) 
        \left ( \begin{array}{c}
                 X_0 \\
                 X_2 
                \end{array} \right ) = {\mathbf{E}}  \left ( \begin{array}{c}
                 X_0 \\
                 X_2 
                \end{array} \right )
                \eea 
where both ${\mathbf{H}}_0$ and ${\mathbf{H}}_2$ are  $N_{\rho} \times N_{\rho} $ diagonal matrices with elements 
 $(H_0)_{n n} = (2 n +|m| + 1 )\hbar\omega_{\rho} + \hbar \omega_z/2 $ and  
 $ (H_2)_{n n}  = (2 n +|m| + 1 )\hbar\omega_{\rho} + 5 \hbar \omega_z/2 $; 
  ${\mathbf{V}}_{0(2)}$ is a $N_{\rho} \times N_{\rho} $ matrix 
 that describes couplings between different radial wavefunctions with $n_z = 0 (2)$. An element of the 
 $N_{\rho} \times N_{\rho} $ matrix ${\mathbf{V}}_{0 2}$ or ${\mathbf{V}}_{2 0 }$ is the cross coupling between one state with $n_z=0$ and 
 another with $n_z =2$. 
 Here ${\mathbf{E}}$ represents a diagonal matrix for the eigenvalues.  

 \begin{figure}
 \centering
 \includegraphics[width=3.5in]{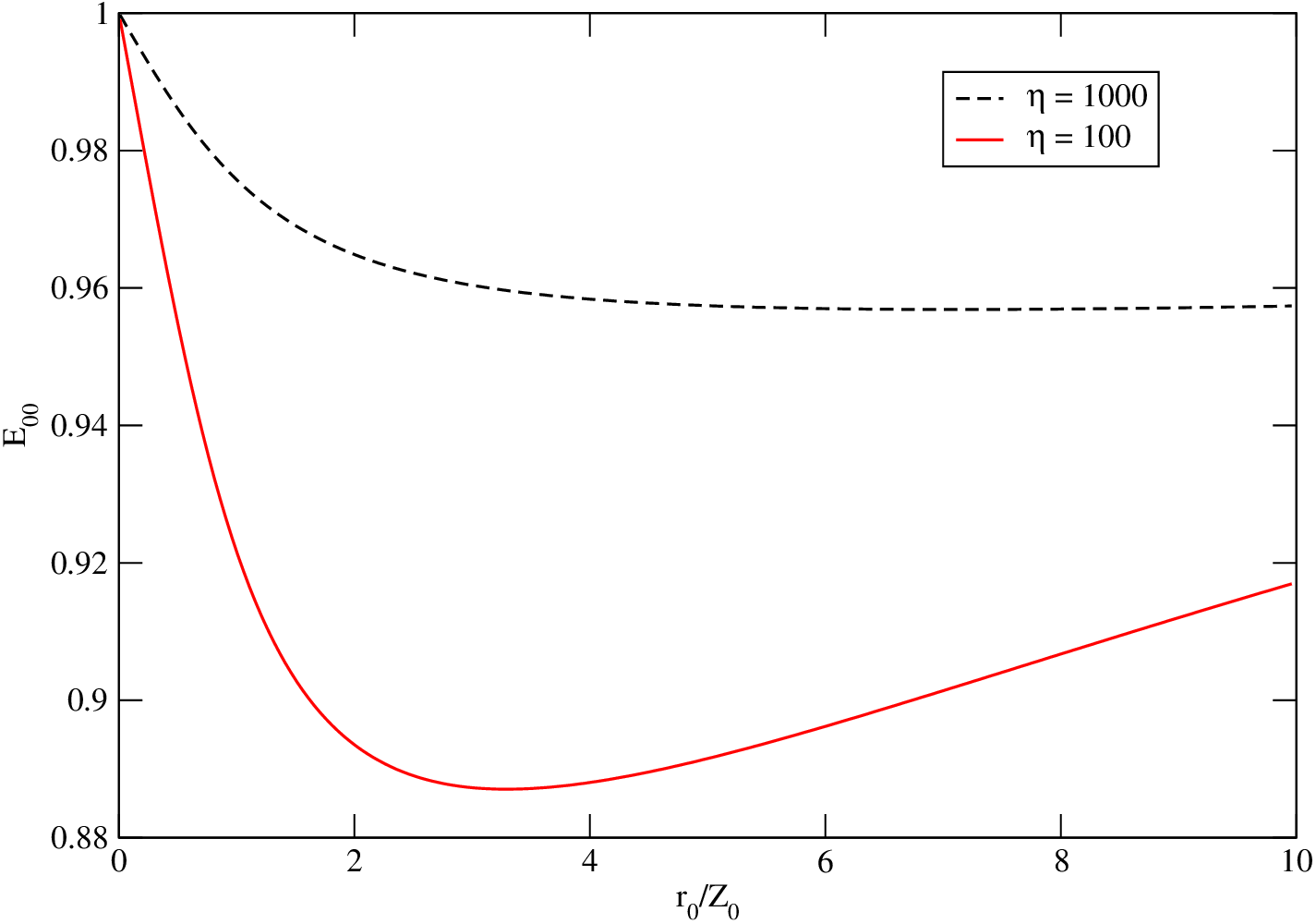}
\caption{Ground state energy ($E_{0,0}$) in unit of $\hbar \omega_{\rho}$  of a pair of  atoms interacting via $V_{\infty}$ 
 is plotted as a function of $r_0$ in unit of $Z_0$  for two different aspect ratios 
$\eta = 100$ (solid  line) and $\eta = 1000$ ( dashed lines).}
 \label{fig:1}
\end{figure}

\section{Results and discussion}

Before we present our results on the effects of $r_0$ and $a_s$ on the bound state properties of two
trapped 
atoms in a quasi-2D or 2D trap ($\eta >\!> 1$), we wish to see how these two parameters affect
the ground-state energy for the isotropic case, that is $\eta = 1$ or equivalently $\omega_x = \omega_y = \omega_z = \omega$ where 
$\omega$ denotes the frequency of the isotropic harmonic trap. 
The results for finite-range 
potential with $a_s = - \infty$ for the isotropic case  
are given in Ref. \cite{ajp:2008:Bhaduri}. We  
we would like to reproduce the known ground-state energy  for the isotropic case  
with our model potentials in the 
limit $a_s \rightarrow \pm \infty$ as a consistency check of 
our numerical method. The eigenstates and eigenvalues of 
two noninteracting atoms in an isotropic trap are well-known and can be characterised by 3 quantum numbers 
which are the principal, orbital angular momentum and and the magnetic quantum numbers.
We calculate the ground-state energy by two methods: One is diagonalisation  of 
the Hamiltonian matrix constructed in the basis of these eigenstates and the other is direct numerical integration 
of Schroedinger equation by Numerov method. We have found that to get convergent value of the ground-state energy 
by diagonalisation method, 
at least 7 lowest basis functions need to be considered for $r_0 > l_0$, where 
$l_0 = \sqrt{\hbar/\mu \omega}$ is  the harmonic oscillator length scale. 
In Table-1, some representative values of the ground-state energies (for different values of $r_0$)  
obtained by diagonalisation with 7 basis functions 
are given and compared with those obtained by Numerov method \cite{numerov,smith}. From this table we 
notice that when $r_0$ is much greater than 
$l_0$,  the two methods yield almost the same results, but when $r_0$ becomes 
comparable to $l_0$, the two results start to deviate considerably. This is because as $r_0$ decreases 
towards or below $l_0$, the two particles start to interact or collide more strongly at or near the trap 
center. As a result, a larger number of harmonic oscillator states can be coupled by the interaction 
necessitating the use of  a larger number of basis functions for convergence of the eigenvalue. We find that the ground-state energy 
for $r_0 > l_0$ and $a_s \rightarrow \pm \infty$ agrees quite well with the results of \cite{ajp:2008:Bhaduri}.  We have checked 
by Numerov method that as $r_0$ decreases below 0.6 $l_0$, the ground-state energy tends to 0.5 $\hbar \omega$ consistent 
with the earlier results for zero-ranged pseudo-potential \cite{foundphys:1998:busch}.  This indicates that 
in the limit $r_0 \rightarrow 0$, our model 
potentials are capable of reproducing  the results of the two atoms interacting with zero-ranged pseudopotential  
in an isotropic harmonic trap.    

We now provide results for quasi-2D trap. 
In  our numerical illustration, in order to  compare readily with the minimum length scale of trapping potential,  we express all length 
scales in unit of $Z_0$. However, all energies are expressed in unit of $\hbar \omega_{\rho}$. We first set $n_z = n_z' = 0$ and 
obtain  bound-state wave functions and bound-state energies by numerically solving Eq. (\ref{psi}). 
The results for the ground state ($\nu =0, m=0$) converge to,  when the matrix dimension  $n_{\rho}$ is equal or greater than 4.  For low lying 
energy eigenvalues $(\nu \le 2, m \le 2)$, we have found convergent results for $n_{\rho} \ge 6$.  
The radial part of the ground state wave function in case of quasi-2D trap are displayed in Figs. 1 and 2 for different positive and 
negative values of scattering length $a_s$, respectively; for the fixed $r_0 = 1 Z_0$. Fig. 1 shows that 
 the ground state amplitude at the trap centre increases with the increase of negative scattering length.  
Exactly opposite effect occurs in the case of  positive scattering length as shown in Fig. 2.  
As can be noticed from these two figures, the interactions affect 
the bound state most significantly near the trap centre rather than near 
the edge of the trap. This 
is due to the fact that at the trap centre trapping potential vanishes. In contrast, at the edge of the trap, trapping potential 
dominates over the atom-atom interactions and consequently the interactions with short or finite range 
have practically no effect on the bound state  near the edge of the trap.

\begin{figure}
 \centering
 \includegraphics[width=3.5in]{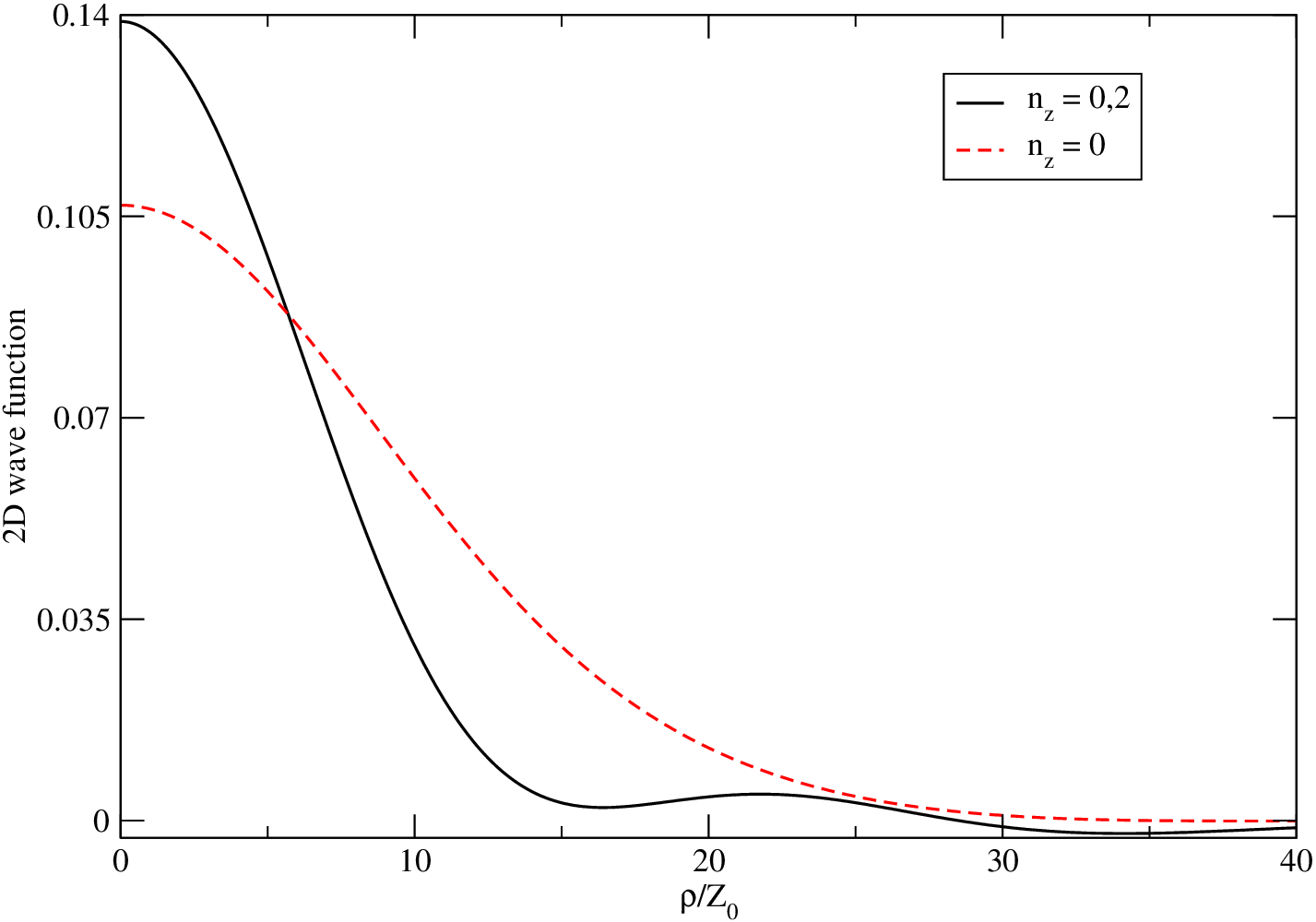}
\caption{ 2D ground state wave function in unit of $\rho^{-1}$ for  
$a_s = \infty $, $r_0 = 1 Z_0$ and $\eta = 100$ considering $n_z = 0,2$ (solid  line) and 
$n_z = 0$ (dashed line).}
 \label{fig3d:1}
\end{figure}

Figures 3(a) and 3(b) exhibit the effects of $r_0$ of $V_{\pm \infty}$ on the 
the radial probability density $P_{2D}(\rho) = |\psi_{00}(\rho)|^2$   at $\rho = 0$. From this figure, 
we observe that as $r_0$ increases  the probability amplitude of relative motion between the two  particles at 
the centre ($\rho=0$) of the 2D or quasi-2D trap increases first up to a certain range $r_0 \sim 3Z_0$ and then decreases. The 
effect is more prominent in case of quasi-2D rather than in 2D case ($\eta$ very large). Note that  $P_{2D}(0)$ is finite at $r_0 = 0$.  

In Fig. 4 we show the effects of the range of the potential $V_{\infty}$ for infinite scattering length on 
the ground-state energy of two-particle bound states in a quasi-2D or  2D harmonic trap.  We notice that when 
$r_0 \rightarrow 0$ the ground state energy is close to unity implying that the interaction  has hardly 
any effect on ground state energy. However, as $r_0$ increases the energy decreases by a few percent. The decrease 
is more prominent if the aspect ratio is relatively smaller. In case of $\eta=1000$, that is $\omega_z = 1000 \omega_{\rho}$, 
the ground-state energy 
decreases  by 
about 4\% as $r_0$ increases from zero to 10$Z_0$,  while in the case $\omega_z = 100 \omega_{\rho}$ the energy decreases by about 
10\% as $r_0$ increases reaching   
a minimum at $r_0 \simeq 3 Z_0$ and then the energy increases as $r_0$ increases past 3$Z_0$. We can infer from these results that 
 the long-range  interactions have  significant 
effects on the ground state energy of a pair of interacting particles in  quasi-2D or 2D traps.  

\begin{table}
\caption{\label{abrefs}Ground state energy eigenvalues $E_{0,0}$ (in unit of $\hbar\omega_{\rho}$) of two-atom  
bound states in a quasi-2D ($\eta=100$) and  2D ($\eta=1000$) harmonic oscillator trap 
 for different values of $a_s$ (in unit of $Z_0$) keeping  $r_0 = 1Z_0$ considering $n_z = 0$ in one case and $n_z = 0,2$ in another case. }
\begin{center}
\begin{tabular}{lcccc} \hline
  $\eta$                                       &$a_s$                  &   $E_{0,0} $  &   $E_{0,0} $   \\ 
  &        & $(n_z = 0)$  & $(n_z = 0, 2)$ \\  \hline 
100                   & 4                 &0.9055   &  0.99996    \\
100                   & 10                 & 0.9153  &0.99997      \\
100                   & $\infty$            &0.9201   &0.99998     \\     \hline 
1000                  &4                 & 0.9715   &0.99999      \\ 
1000                  &10                 & 0.9743   &0.99999     \\ 
1000                  &$\infty$             &0.9757    &0.99999   \\     \hline
\end{tabular}
\end{center}
\end{table}

In Table 2 we display a few low lying energy eigenvalues with magnetic quantum numbers  $m=0$ and $m=1$ 
of eigenstates of two interacting particles 
in a quasi-2D ($\eta=100$) and 2D ($\eta=1000$) traps for different values of $a_s$ but for fixed 
$r_0=1 Z_0$. From this table several inferences can be drawn. First,  the deviation of the ground-state energy 
 in the interacting case with positive $a_s$ from that 
in the noninteracting case  ($1 \hbar \omega_{\rho}$) 
is more prominent compared to the similar case with negative $a_s$. In contrast, as we have observed from Figs 1 and 2, that 
the modification in ground-state wave function at or near the  centre of the 2D trap ($\rho=0$) due to an increased  negative 
scattering length is larger than that due to an  increase  in positive scattering length by the same amount. This means that 
an attractive interaction i.e. negative scattering length has more prominent effect on the central probability density  than on the 
ground-state energy, while the reverse is true for repulsive interaction i.e. positive scattering length. Second, 
as the positive $a_s$ increases, the energies 
increase while for increasing negative $a_s$ the  energies decrease. For both the limits $a_s \rightarrow + \infty$ and 
$a_s \rightarrow - \infty$, the energies approach to the same value. Third, the decrease of energies with 
lowering of $\eta$,  as can be noticed from this table,   can be attributed to the deviation from 2D nature. Fourth, the change in $a_s$ alters 
the ground state energies  more prominently  rather than the excited state energies implying that 
the interactions  have the most significant effects on the ground state.  
Fifth, this table also shows that the interactions have quite small influence on the bound states with  $m=1$. 
In contrast, the interactions have pronounced effects on bound states  with $m = 0$. 
Since 2D harmonic oscillator wave functions for $m \neq 0$ go to zero as $\rho \rightarrow 0$, 
short range interactions between two atoms can hardly affect those oscillator states.

Now, we discuss how good are the conditions of large $\eta$ that we have used to describe quasi-2D or 2D regime of the traps. We numerically solve
Eq. (16) to know how significant are the effects of axial excitation with $n_z = 2$ on the ground-state energy and wave function. 
In Table 3 we compare ground-state energies for the two cases $n_z=0$ and $n_z = 0$ and 2 with $\eta = 100$ and $\eta = 1000$ for 3  
values of $a_s$. We notice that 
for $\eta = 100$ the ground-state energy for $n_z = 0$ and 2  cases deviates by more than 7\% from that for $n_z=0$ case, while the corresponding 
deviation in case of $\eta = 1000$ is about 2\%. We display the effects of this axial excitation on the ground-state wave function 
in Figs. 5 and 6 for $\eta=100$ and $\eta=1000$, respectively. We observe that in both the cases the deviation of the wave 
function is small near the trap 
centre and near the edges, while in the intermediate radial separations the deviation for $\eta = 100$ is quite significant,  but for 
for $\eta = 1000$ the deviation is small and may be ignored for all practical purposes. From this analysis, we may infer that while 
for $\eta \le 100$ one can not ignore axial excitations and has to consider full 3D picture, in case  $\eta \ge 1000$ the effects 
of axial excitations on the ground-state properties are quite small.  For $  \eta >\!> 100$, one may describe a quasi-2D regime 
where one can approximately calculate the ground state eigenenergy by ignoring the effects of axial excitations.

\begin{figure}
 \centering
 \includegraphics[width=3.5in]{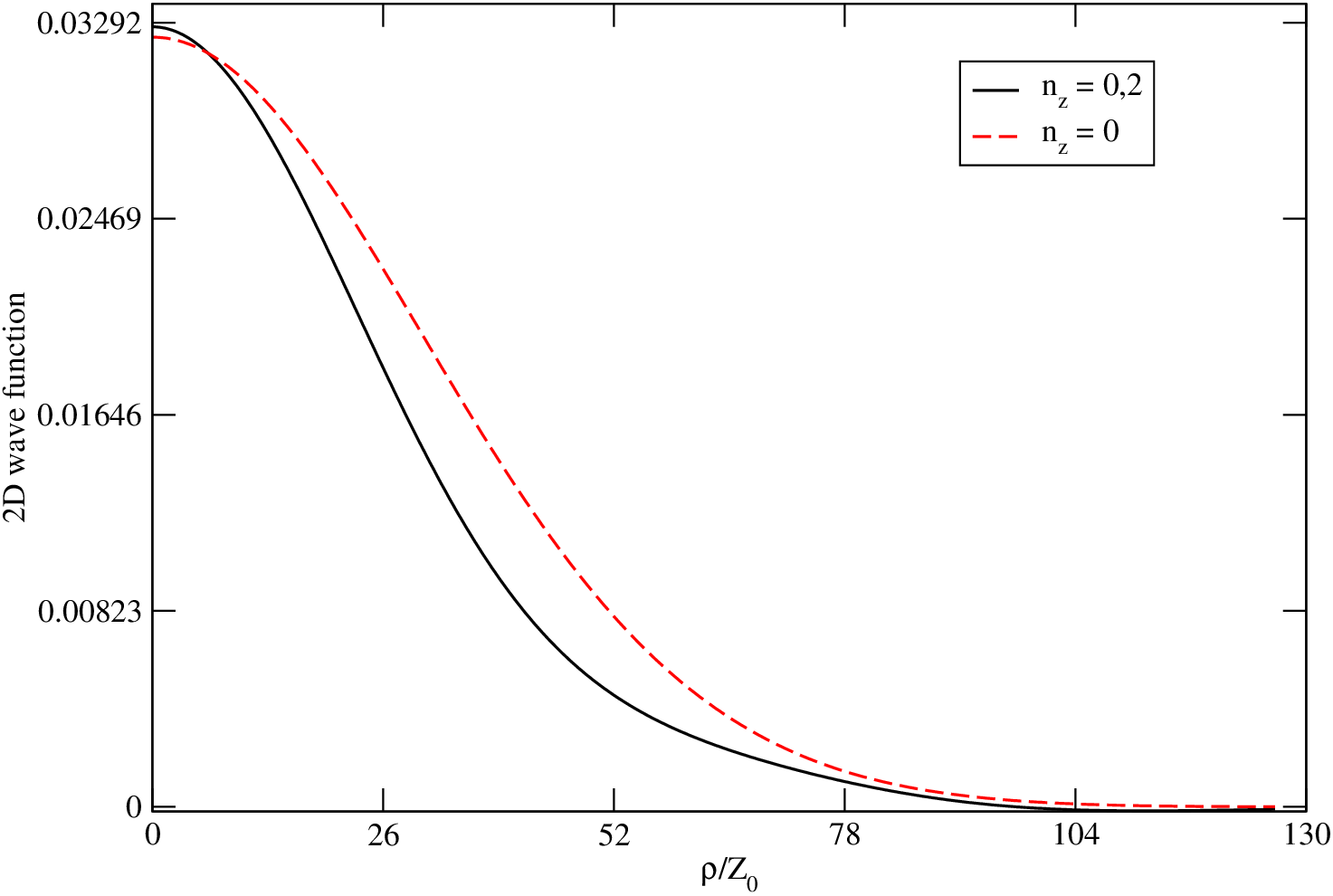}
\caption{ 2D ground state wave function in unit of $\rho^{-1}$ for  $a_s = \infty $, $r_0 = 1 Z_0$ and $\eta = 1000$ considering $n_z = 0,2$ (solid line) and $n_z = 0$ 
(dashed line).}
 \label{fig:1}
\end{figure}

\section{conclusions}

In conclusion we have presented the results of a microscopic model study on the effects of the range 
and the scattering length of a class of two-parameter  model potentials on the  properties of trapping bound-state
of two atoms in a quasi-2D trap.  As discussed in the introduction part,
the range $r_0$ of atom-atom interaction in the pseudo-potential approximation
is usually neglected. However, when $a_s \rightarrow \pm \infty$ which is identified as ``unitarity limit'' in current 
cold atom literature,  the only length scale available for 
expansion given in Eq. (\ref{expan}) is the effective range $r_0$. 
In this limit,  the range of interaction and the momentum 
dependence of $s$-wave scattering amplitude may not be negligible. Finite range of two-body interactions is also important 
in the context of dipolar systems such as magnetic dipolar atoms or electric polar molecules which have long-range and anisotropic 
interactions. Recently, physics of cold polar molecules have attracted tremendous research interests because of the long-range
nature of their intermolecular interactions. Furthermore, ultracold polar molecules in 2D trap may serve as an intriguing system 
for exploring new physics with anisotropic interactions. 
In view of these recent developments in the frontier areas of ultracold atoms 
and molecules, the results of this study  may be useful to develop an understanding of 
microscopic picture about those finite-ranged interactions in 2D trap the counterpart of which in 3D corresponds to 
the tunable Fano-Feshbach resonances.

\vspace{0.5cm} 

\noindent 
{\bf Acknowledgements} \\
PG and AR are thankful to CSIR, Govt. of India, for support. 

\vspace{0.05in}


\end{document}